\documentclass[preprint]{elsarticle}
\usepackage{graphicx,amssymb}

\begin{document}
\begin{frontmatter}

\title{Structural differences between open and direct communication in an online community}

\author[icelab]{Fariba Karimi}
\author[donders]{Ver\'onica C. Ramenzoni }
\author[icelab,skku,socsu]{Petter Holme}

\address[icelab]{IceLab, Department of Physics, Ume{\aa} University, 90187 Ume\aa, Sweden}
\address[donders]{Donders Institute for Brain, Cognition, and Behavior, Radboud University, The Netherlands}
\address[skku]{Department of Energy Science, Sungkyunkwan University, Suwon 440-746, Korea}
\address[socsu]{Department of Sociology, Stockholm University, 10691 Stockholm, Sweden}

\begin{abstract}
Most research of online communication focuses on modes of communication that are either open (like forums, bulletin boards, Twitter, etc.) or direct (like e-mails). In this work, we study a dataset that has both types of communication channels. We relate our findings to theories of social organization and human dynamics. The data comprises 36,492 users of a movie discussion community. Our results show that there are differences in the way users communicate in the two channels that are reflected in the shape of degree- and interevent time distributions. The open communication that is designed to facilitate conversations with any member, shows a broader degree distribution and more of the triangles in the network are primarily formed in this mode of communication. The direct channel is presumably preferred by closer communication and the response time in dialogues is shorter. On a more coarse-grained level, there are common patterns in the two networks. The differences and overlaps between communication networks, thus, provide a unique window into how social and structural aspects of communication establish and evolve.
\end{abstract}

\end{frontmatter}

\section{Introduction}

Online media have created unprecedented means of communication and with them came new ways of organizing many aspects of our lives. One of the advantages of the social media revolution is the possibility to record and study information about interpersonal communication much more easily. Researchers have explored different aspects of online communication, such as the timing of communication, the kind of interactions, and the type of networks of interactions that emerge in online communities. Two kinds of communicational interactions have received special attention in the literature: direct communication, that is mostly one-to-one e.g.\ e-mails~\cite{danowski_edison-swift,bornholdt:email,mejn:email,tyler:email,eckmann:dialog}, Internet dating~\cite{pok}, gaming communities~\cite{szell_online_game}, social media~\cite{traud_facebook,Ahn_online_ntw,hogan_social_ntw, Konstas:lastfm}, and mobile phone communication~\cite{onnela_etal,eagle_phone,ratti_mobile_landscape,blondel_phone,Hidalgo_mobile_persistance}â and open communication, that is mostly one-to-many such as bulletin boards, Twitter, photo-sharing services, etc.~\cite{Mislove:2007,Adamic-yahoo-forums,Chun-guest_book}. Distinct studies of these two types of communication have demonstrated that many aspects of human online communication are universal. Communication networks have a broad, power-law like distributions of degree (the number of neighbors in the network)~\cite{bornholdt:email,mejn:email,pok} and the interevent time distributions of individual activities (like the sending of e-mails) is also fat-tailed~\cite{johansen:resptimes}. From these observations, researchers have derived models that seek to explain the evolution of different forms of online communication and their temporal correlations~\cite{Mislove:2007,barabasi:bursty,HSrev}. 

However, few studies have compared different types of communication to study their structure beyond the broad-brush-stroke universal patterns mentioned above~\cite{szell_online_game}. This is not due to lack of interest in the issue but rather to the difficulty in accessing data corpuses that contain simultaneous information about different types of communication. Such a corpus requires \emph{multiplex} network data, that covers the same individuals who engage in tow or more types of communications over a prolonged period of time. This paper aims at providing a comprehensive, comparative study of the networks of direct and open communication for a community of individuals. We studied a corpus of communications from a large population of individuals obtained over the course of seven years.

The online community we explored is composed of individuals that engage in discussions about their film interests. It offers users two exclusive means of communication. One of them is a simple user-to-user messaging channel---we refer to it as \emph{Messages}---where each user can send text messages to another user privately and only one user at a time.  The other channel is an open forum (we denote it by \emph{Forum})---a visible platform where users can discuss with other users, as many as are willing to participate. Posts in a topical forum thread are, except the first one in a thread, always in a response to one of the previous posts. As a consequence, \emph{Forum} is similar to \emph{Messages} in that it facilitates a dialogue between users, but differs from \emph{Messages} in that they are visible to the rest of the community. These differences and overlaps between means of communication reflect differences in how the communication channels are organized---by the platform or by users---in both communities~\cite{Stivers_turn_taking}. For instance, one of the socially relevant differences between the platforms is that they require different degrees of collaborative effort~\cite{clark:using_language} such that in \emph{Messages}, the emergence of relationships between users imposes a higher degree of demands on each individual, while in \emph{Forum} the requirement to collaborate is distributed across all active users. 

In this paper, we seek to explore characteristics of the individuals and networks in terms of their behavior and structure, respectively. We will try to explain the characteristics in terms of the restrictions and capabilities of different modes of communication. We expect to draw conclusions about how the differences in visibility alter the social processes behind the two types of communication and how similarities in their communicative behavior help unveil aspects that are universal to online communication regardless of the platform in which they are instantiated.

\section{Rationale and Hypotheses}

In order to address the nature of these two communication networks, we took an exploratory strategy. First, we studied the behavior and characteristics of individual users such as the in-degree and out-degree of the individuals, the level of social involvement, interactivity and interevent time.  Second, we explored the network  structures in terms of reciprocity, transitivity and assortativity. Last, we explored the imprints of social processes,  such as the Jaccard similarity of neighboring contacts and triangle motifs. In terms of the behavior of individual users and the kind of sociality promoted by each communication platform, we made a number of hypotheses. We expected that \emph{Forum} would allow individual users to send out and receive more communication---here we call it \emph{contact}---compared to the \emph{Messages} network. These hypotheses would both be in line with the regulations of the site and consistent with each mode of communication serving their particular purpose. To seize the behavioral aspects, we introduced \emph{sociality} and \emph{interactivity} measure. Sociality captures to what extent people are open to interact with strangers and interactivity captures overall activity level of individuals. We expected sociality and interactivity to be reflected from the mode of communication. Additionally, we expected the differences in forms of sociality might also lead to differences in the timing of the interaction and the general involvement of individuals.
 At a community level,  we expected the relationships between individuals to be structured in different ways depending on the amount of commitment to the community and other individuals  elicited by the respective mode of communication. Because of its resemblance to dyadic conversations, we anticipated users would show a higher degree of reciprocity in \emph{Messages}  compared to \emph{Forum}, where posts might not require a commitment from other \emph{Forum} users to respond to comments.  However, we expected both \emph{Forum} and  \emph{Messages} platforms would give rise to triangles in the interaction networks---a signature of offline social networks~\cite{newman_social_netwrok}. To study this point,  we looked at the amount and characteristics of triadic relationships each mode of communication gave rise to. For both online communication networks we expected triadic relationships to be more frequent than what one would expect by mere chance encounters. We classified triadic interactions and looked at whether different triadic interactions emerged from each platform. Finally, we looked at the Jaccard similarity index for both networks; here we anticipated that the structure of relationships between users to some extent overlap between the networks.

\section{Datasets and basic properties}

\begin{table}
\caption{Basic statistics of the datasets}
\begin{tabular}{lr}
\hline\hline
Number of users in \emph{Forum} & 6,269\tabularnewline
Total number of contacts in \emph{Forum} & 1,357,535\tabularnewline
Number of  edges (interacting pairs) in \emph{Forum} & 122,369\tabularnewline
Number of users in \emph{Messages} & 35,564\tabularnewline
Total number of contacts in \emph{Messages} & 490,442\tabularnewline
Number of edges in \emph{Messages} & 94,655\tabularnewline
Number of users in \emph{Forum}  and \emph{Messages}  & 5,341\tabularnewline
Total sampling time & 6.9 years\tabularnewline
\hline\hline
\end{tabular}
\label{tab:Filmtipset.se-in-statistics} 
\end{table}

The datasets comes from an online community of movie enthusiasts (also described in Ref.~\cite{filmtipset_german}). It consists of communications between 36,492 users for a period of about 7 years. The data does not contain any demographic information or the content of the messages. To join the community, users need to register and set up a profile. Even though it is discouraged and actively counteracted, it is possible that the same individual register several times. There are some discussion on the \emph{Forum} about possible multiple identities, but we believe this is a rather rare phenomenon. Profile pages consist of statistics of the users, such as the level of online activity in \emph{Forum} and \emph{Messages}, users with similar movie tastes, list of friends and lists of rated movies. The profile page also contains an interface that allows users to access the \emph{Messages} channel. \emph{Forum} is organized into topical categories (films and film industry, user posted lists, games and polls, non-film related, etc.), which have changed over the sampling period when new categories were created. In each category, users can post threads on specific topics that often grow into discussions of a hundred or more users. The number of posts in a \emph{Forum} thread can vary from 1 to 1421. 

All identifying information was removed from the dataset before we received it.  None of the analysis that are described here focused on single individuals, all statistics are aggregated over the entire population. To study the network topology,  we aggregate contacts in a multiplex static network where an edge means that two vertices have been in contact at some time.

We obtained two datasets; each of them consists of all time-stamped contacts between users. The two datasets, were composed by lists of triples $(i, j, t)$, or \emph{contacts}, where $i$ is the sender of a \emph{Messages} or \emph{Forum} contact, $j$ is the recipient, and $t$ is the time. The forum pages in this community are constructed such that each comment -- except the first one in the thread -- has a link to a previous comment in the corresponding thread and a time stamp. The tree-like structure of the \emph{Forum} conversations enables us to establish a link between two users if they comment each other in the \emph{Forum} dataset.

In the \emph{Forum}, users create various topics and participate in different discussions. Users are allowed to reply to their own comments. We can represent such multiple answers from the same user as loops in the network. In the \emph{Forum} network, there are 1755 loops during the entire sampling time.
\emph{Messages} works as an online personal space for individuals~\cite{ananda_cyberspace}. The users have more control of their person-to-person communications and information sharing. In contrast, \emph{Forum} is mainly designed for socializing, meeting new users and discussing similar interests~\cite{harris_secondlife}. 
There are 7,183  edges present in both \emph{Forum} and \emph{Messages}. This is $3.6\% $ of the total number of vertex pairs connected by any edge.  Among these $3.6\%$, $67\%$ of their first contacts happened in  \emph{Forum}. Table~\ref{tab:Filmtipset.se-in-statistics} lists basic statistics of the data.

\section{Results}

We first describe the structural differences and similarities between the communication channels at an individual level, then we describe the structure at the level of pairs, or edges, and last we describe system-wide properties. Note that because of limited access to the demographic differences in the population level, we focus on the measures that can be discussed beyond those differences.

\subsection{Individual level}

\begin{figure}
\includegraphics[width=1\linewidth]{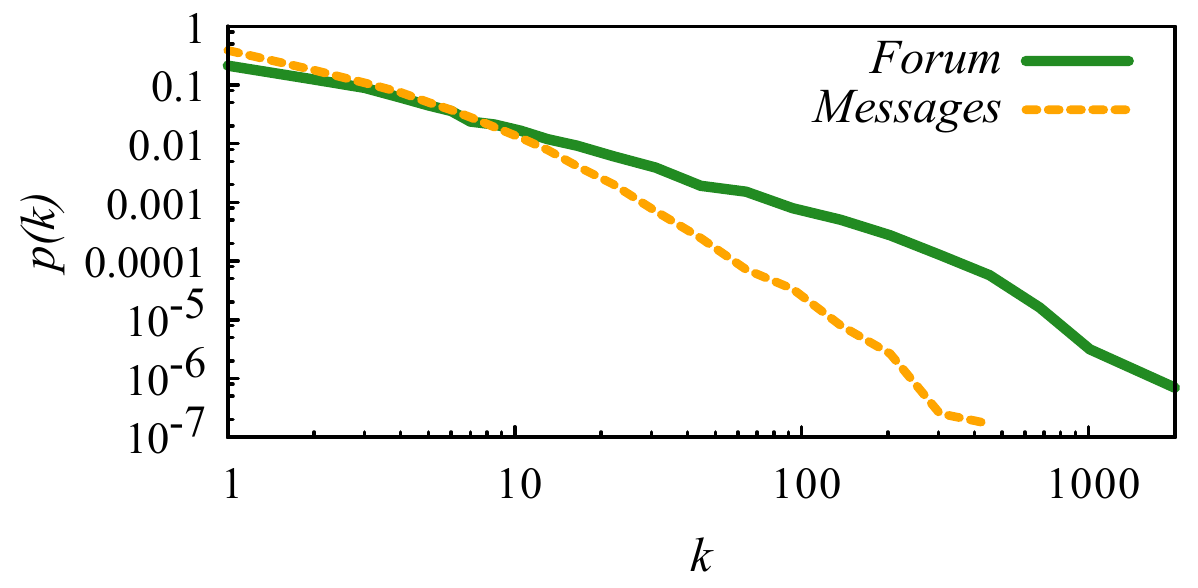}
\caption{\textbf{In-degree distributions of \emph{Forum} and \emph{Messages}.} The in-degree distribution in \emph{Forum} is broader than in \emph{Messages}. The data is log-binned.}
\label{fig:in_deg_dist}
\end{figure}

\begin{figure}
\includegraphics[width=1\linewidth]{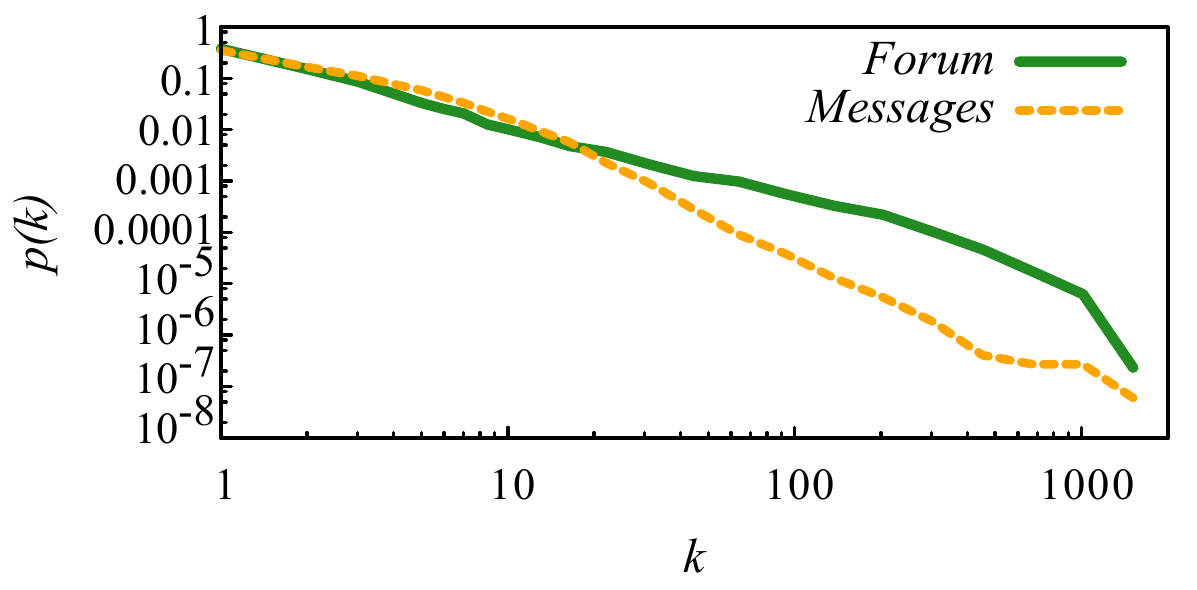}
\caption{\textbf{Out-degree  distributions of \emph{Forum} and \emph{Messages}.} The out--degree distribution in \emph{Forum} is broader than in \emph{Messages}. The broader distribution in \emph{Forum} shows that the open space enhances broader degree of communication. The data is log-binned.}
\label{fig:out_deg_dist}
\end{figure}

\paragraph{Degree distribution}
Figs.~\ref{fig:in_deg_dist} and \ref{fig:out_deg_dist}, compare in- and out-degree distribution in \emph{Forum} and \emph{Messages}. On average, the degree distribution is broader  in \emph{Forum} compared to \emph{Messages}. In addition, \emph{Messages} showed a slightly higher number of users with low degree of connectivity. As we expected, \emph{Forum} allows for users to interact with a larger number of users, whereas in \emph{Messages} users have fewer and selective interactions with others. These differences indicate that both means of communication have different social functions. The degree distribution of an undirected network looks similar to the in- and out-degree distributions and therefore omitted.

\begin{figure} 
\includegraphics[width=1\linewidth]{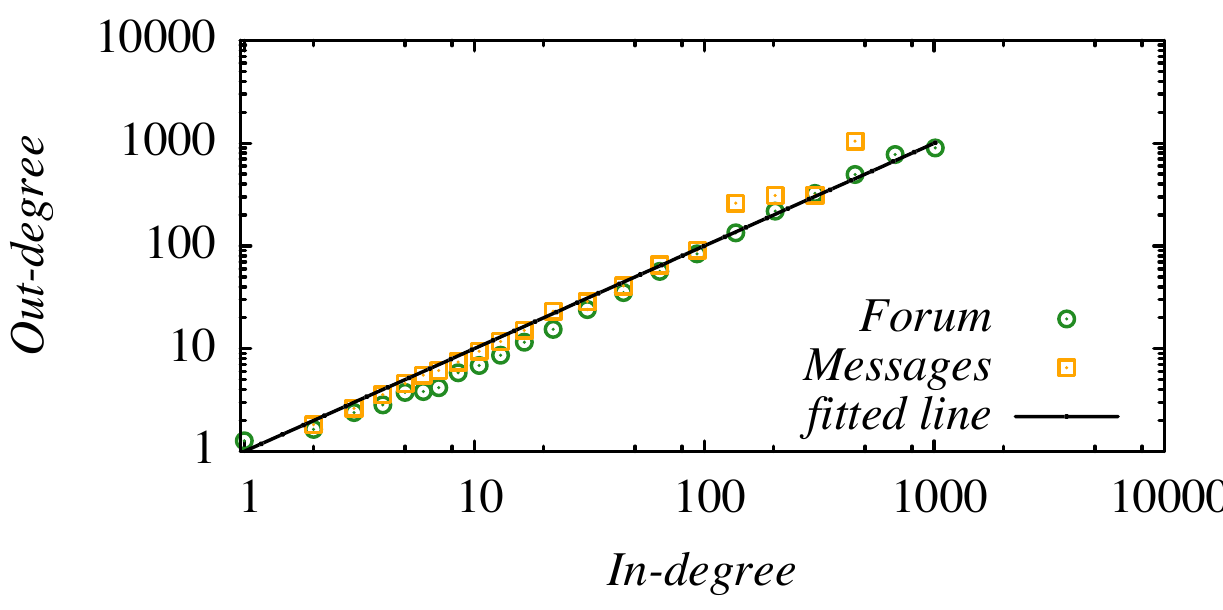}
\caption{\label{fig:in_out_activity}\textbf{Comparing in-degree and out-degree of \emph{Forum} and \emph{Messages}.} The scales are logarithmic. At the level of individuals, activity of sending and receiving \emph{Messages} and posting \emph{Forum} are correlated. For the higher out-degree, users are sending out more messages than receiving messages.}
\end{figure}

However, the analysis of the in- and out-degree distributions for each individual, showed that in- and out-degree were positively correlated (see Fig.~\ref{fig:in_out_activity}). This was true for both networks---users who post more \emph{Forum} comments or send more \emph{Messages} are likely to receive more comments from the other users of the community. For users present in both \emph{Forum} and \emph{Messages}, there is a weak correlation between the out-degrees in the respective channels (not shown). This means that these multichannel users do not segregate into those with a preference for one specific channel, rather they can be divided into more or less active ones.

\begin{figure} 
\includegraphics[width=1\linewidth]{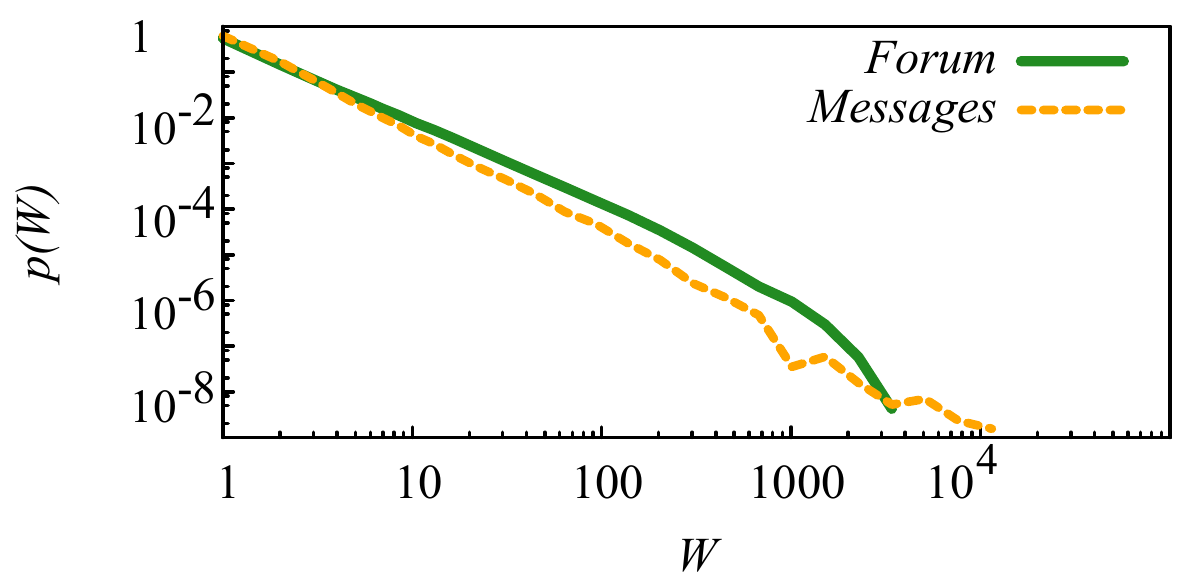}
\caption{\label{fig:link_weights}\textbf{Link weight distribution in \emph{Forum} and \emph{Messages}.} The weight distribution is broader in \emph{Messages}, reflects the fact that direct communication has higher rate of reciprocity and exchange.}
\end{figure}

Another way of looking at users' interactions is to measure the weight of interaction. We do this by defining an edge weight between two vertices $i$ and $j$, $w_{ij}$. The edge weight corresponds to the number of times that the vertices interact. From the Fig.~\ref{fig:link_weights} we see that the weight distribution is broader in \emph{Messages}. This reflects the fact that more one-on-one interaction happen in direct communication.

\paragraph{Interevent time}

\begin{figure}
\includegraphics[width=1\linewidth]{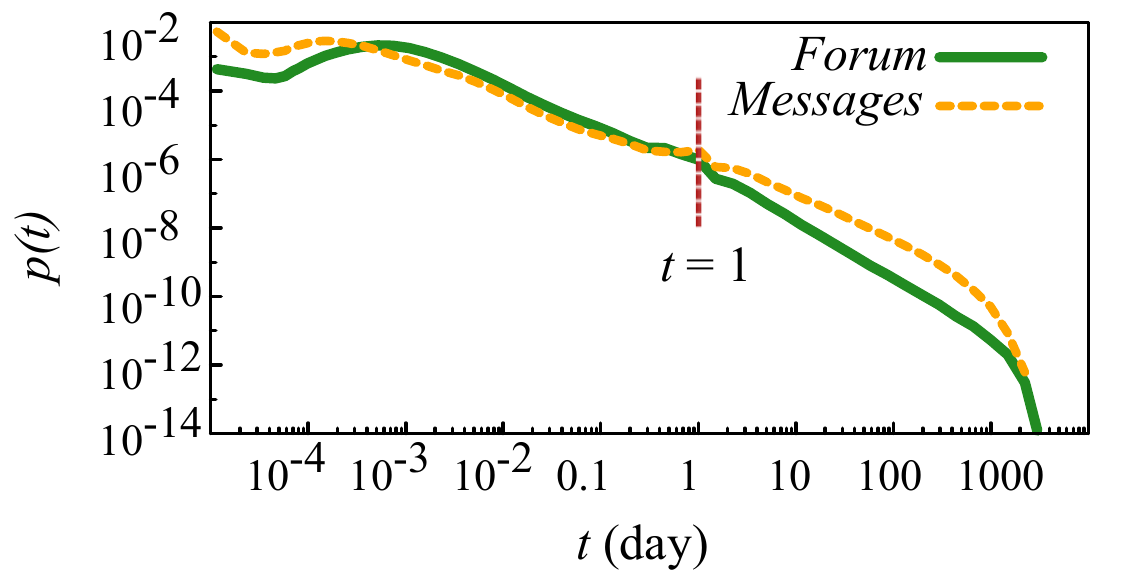}
\caption{\textbf{\label{fig:Interevent-time-comparing}Interevent time distributions of \emph{Forum} and \emph{Messages}.} The distribution shows the time interval performing two  consecutive posts in \emph{Forum} and \emph{Messages}.  The data is log-binned.}
\end{figure}

In order to capture the dynamics of individual communication over time, we looked at the time between successive activities of individuals (i.e., the time individuals take to send or reply to \emph{Messages} and \emph{Forum}). While interevent time analysis have been done before~\cite{pok, malmgren_interevent}, there are only few previous studies that compare interevent time statistics by means of communication e.g. phone calls and SMS~\cite{karsai2012universal}. In this paper, we investigate whether interevent times obtained from one population change depending on the means of communication users use. We calculate the interevent time as the time between two consecutive posts or messages by the same user. The interevent time is shown in Fig.~\ref{fig:Interevent-time-comparing}. In general, interevent times showed differences in the time users take to reply to each other in both networks. While logged in (i.e., for small time windows), users are slightly faster in replying to \emph{Messages} than \emph{Forum} posts, which suggests that users engage in an ongoing conversations using \emph{Messages}.  For time windows larger than 24hs, users are less likely to respond to \emph{Forum} contacts than \emph{Messages}. The distribution of interevent time suggests that in a forum discussion, users' response times are  based on ongoing discussions that are reinforced by other users, in other words the fact that many people can participate in a forum discussion strengthen the feedback of a discussion and consequently the timing of the discussions. Such dynamics do not exist in one-on-one messaging and users are more likely to respond messages longer than a day.

\paragraph{User sociality and interactivity}

\begin{figure}
\includegraphics[width=1\linewidth]{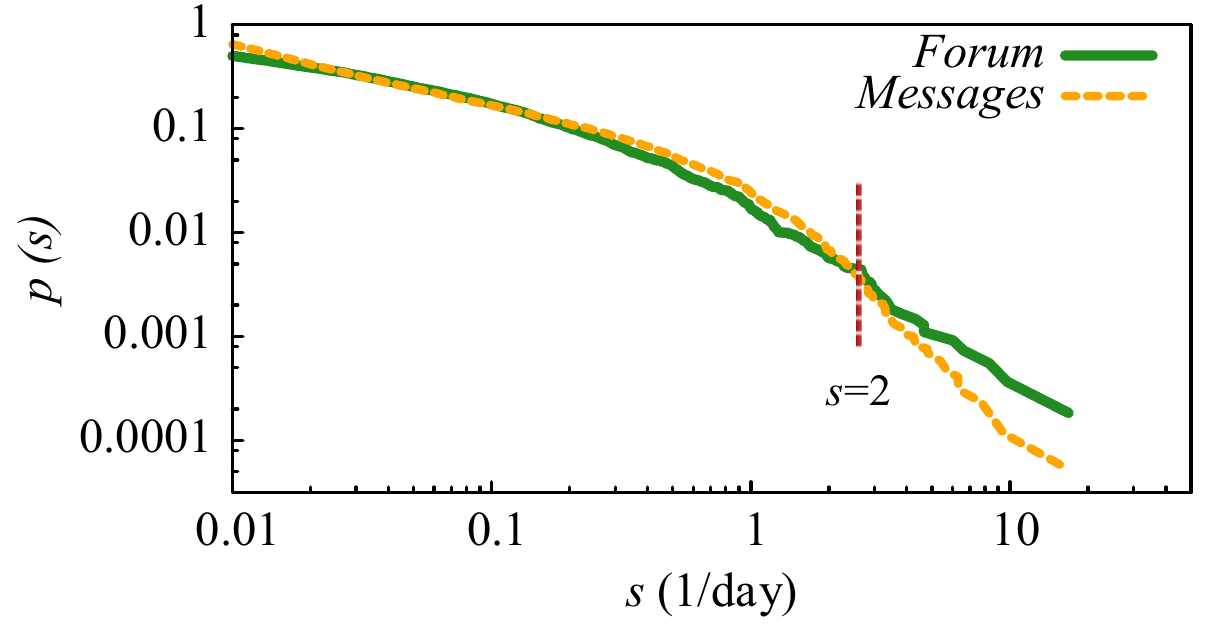}
\caption{\textbf{Cumulative distribution functions of users' sociality in
\emph{Forum} and \emph{Messages}.} The sociality is calculated by normalizing degree of connection to the users lifetime in days. There is broader distribution of sociality in \emph{Forum} for $s>2$. The scales are logarithmic. }
\label{fig:vertex_sociality}
\end{figure}

We introduced sociality as the total number of unique edges that a user established, divided by the user's total lifetime in the data. Here, to get unbiased results, we discard users whose presence in the datasets is less than a day that is 30\% of the total users.
Let $k_{i}$ denotes the degree of user $i$, i.e.\ the number of other users $i$ has interacted with. The time interval between the first appearance of $i$ in the dataset and the last appearance, is denoted by $\tau_{i}$ that it is shown in the scale of days. To calculate the time interval, we compare users first and last appearance time in both datasets and choose the earliest and the latest time that the user appeared in either of the platforms. Therefore the sociality of the user $i$, $s_{i}$ equals to
\begin{equation}
s_{i}=\frac{k_{i}}{\tau_{i}}
\end{equation}

Fig.~\ref{fig:vertex_sociality} shows users' sociality in \emph{Forum} and \emph{Messages}. We observe an upper boundary ($s=18$) for sociality of users in both datasets. That means in maximum people communicate with 18 people per day. While for low and medium sociality, \emph{Forum} and \emph{Messages} are similar, for higher sociality ($s>2$), people are more social in \emph{Forum} than \emph{Messages}. We conclude that the sociality seems independent of the mode of communication.

\begin{figure}
\includegraphics[width=1\linewidth]{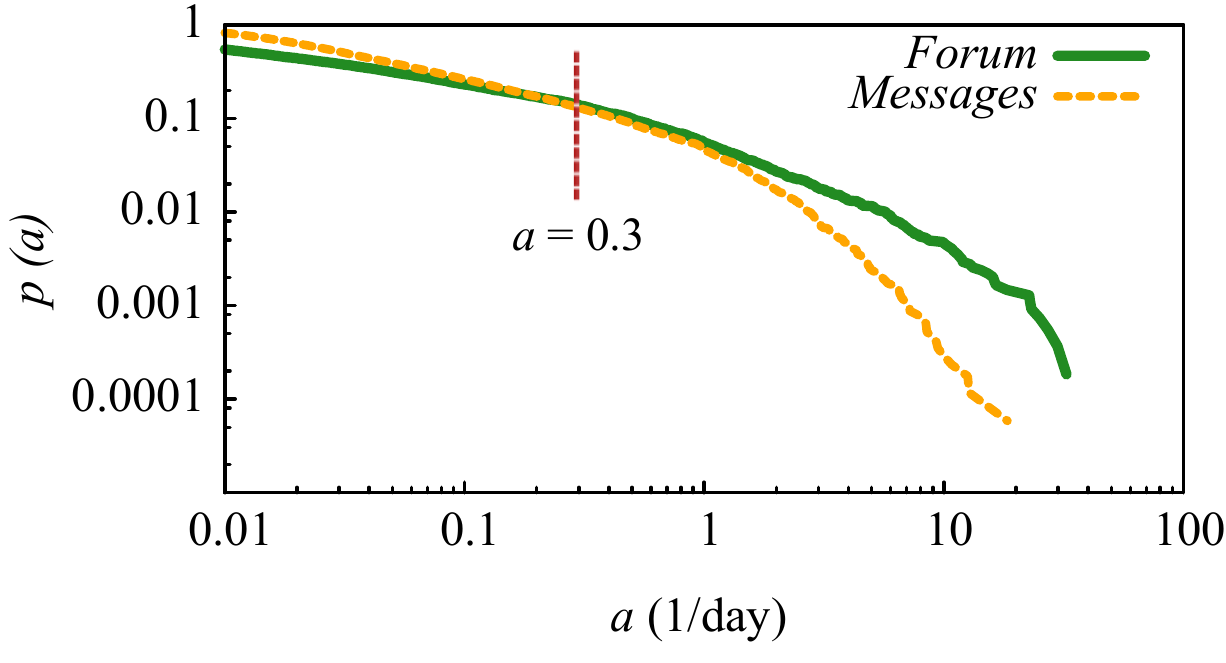}
\caption{\textbf{Cumulative distribution functions of users' interactivity in
\emph{Forum} and \emph{Messages}.}  The plot shows the total number
of \emph{Messages} or \emph{Forum} posts a user has sent during their time in  the community.  There is broader distribution of interactivity in \emph{Forum} for $a>0.3$. The scales are logarithmic.} 
\label{fig:messaging_capacity}
\end{figure}

Now we turn to explore the user \emph{interactivity}---the rate of activity of a user. Interactivity $a_{i}$ is measured as the number of \emph{Messages} or \emph{Forum} posts a user sent out, divided by the total time that the user was active in the community. 
Let $\pi_{i}$ denote the total \emph{Forum} posts or \emph{Messages} a user $i$ sent to the community. The time interval between the first appearance of $i$ in the dataset (either \emph{Messages} or \emph{Forum}) and the last appearance, is denoted by $\tau_{i}$. Therefore, the interactivity of the user $i$ is
\begin{equation}
a_{i}=\frac{\pi_{i}}{\tau_{i}}
\end{equation}

Fig.~\ref{fig:messaging_capacity} shows the distribution of interactivity of all users comparing the two communities. 

From the figure, we observe an upper bound in both networks. This bound is larger and the distribution is broader in \emph{Forum} $(a_{max}=32)$ than in \emph{Messages} $(a_{max}=18)$. There is a cross-over $(a=0.3)$ where higher interactivity starts increasing in \emph{Forum}. The \emph{Forum} platform is designed for promoting conversations and interactions and it is reflected from broader distribution of interactivity in the Fig.~\ref{fig:messaging_capacity}.

\begin{figure}
\includegraphics[width=1\linewidth]{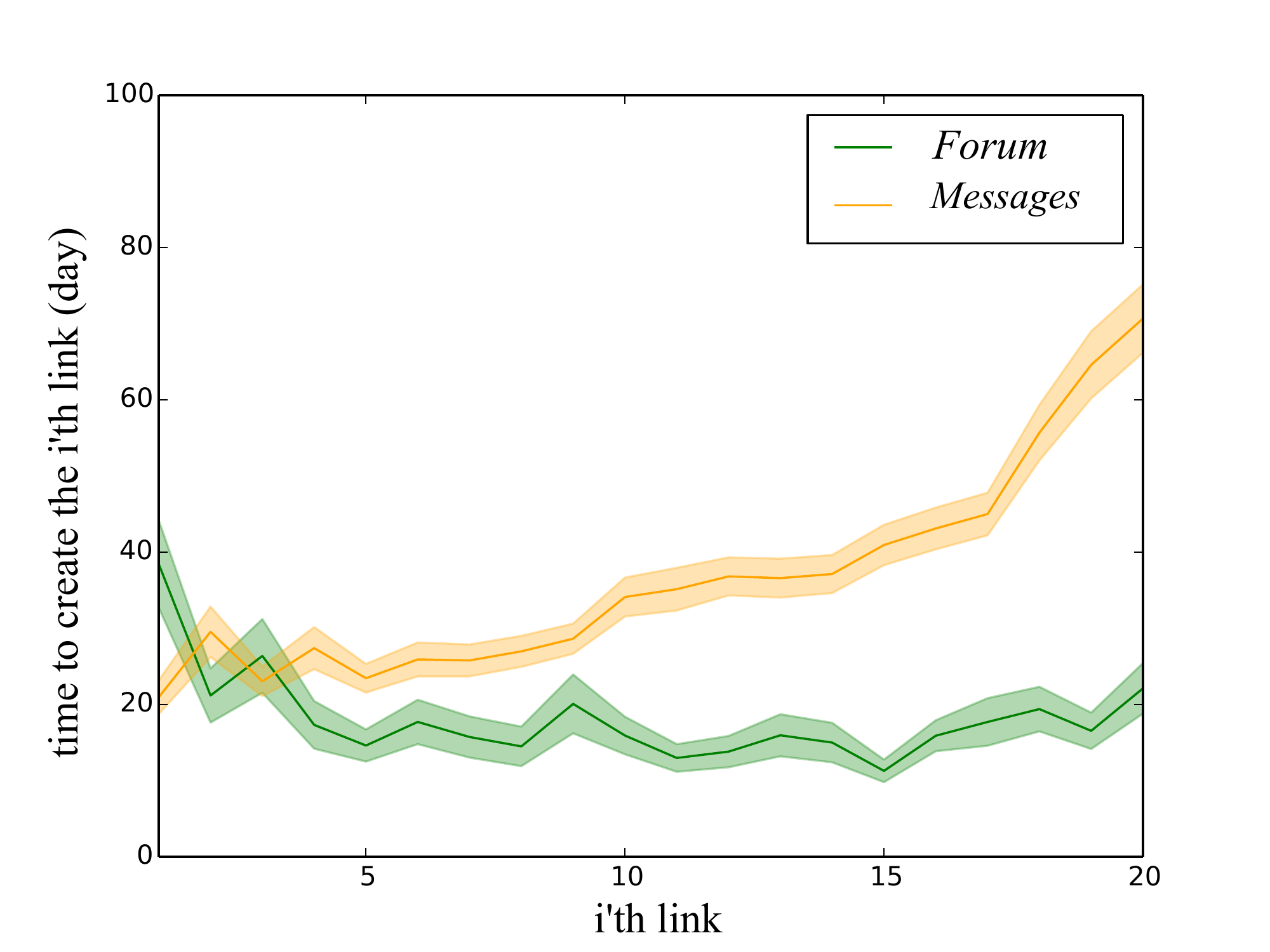}
\caption{\textbf{Average time that is needed by the users to create the $i$'th edge.} In \emph{Forum} the expected time to create new edge decreases with number of edges and eventually plateaus. However in \emph{Messages}, the time increases with number of added edges, reflecting that creating edges in \emph{Messages} demands larger social investment.} 
\label{fig:link_careation}
\end{figure}

Sociality captures the capacity of users to make new edges during their lifetime in the platform. Yet, it is intuitive to measure the average time between making the $i$'th and the $(i+1)$th edge for users in the two online spaces. To measure how long it takes to make new edge, we select users with 20 edges or more in \emph{Forum} than \emph{Messages}  and we measure the time between creating $i$'th and the $(i+1)$th edge. The results are shown in Fig.~\ref{fig:link_careation}. In general, we observe different trends between the two online spaces. The results suggest that, as we speculated before, the creation of edges in \emph{Messages} demands more social investment.

Moreover, for users that are active in both \emph{Messages} and \emph{Forum} there is a strong correlation between interactivity in the respective channels. Users that are presented in both datasets are 3\% of the population, that means there are 15\% of users in \emph{Forum} that are not present in \emph{Messages} and 82\% of users who send \emph{Messages} that never involve themselves in \emph{Forum}. On the other hand, there are 15\% of users in \emph{Forum} that are not present in \emph{Messages} and 80\% of users who send \emph{Messages} that never involve themselves in \emph{Forum}.

\paragraph{Effect of the first message contact on dual-channel pair communication}

\begin{figure}
\includegraphics[width=1\linewidth]{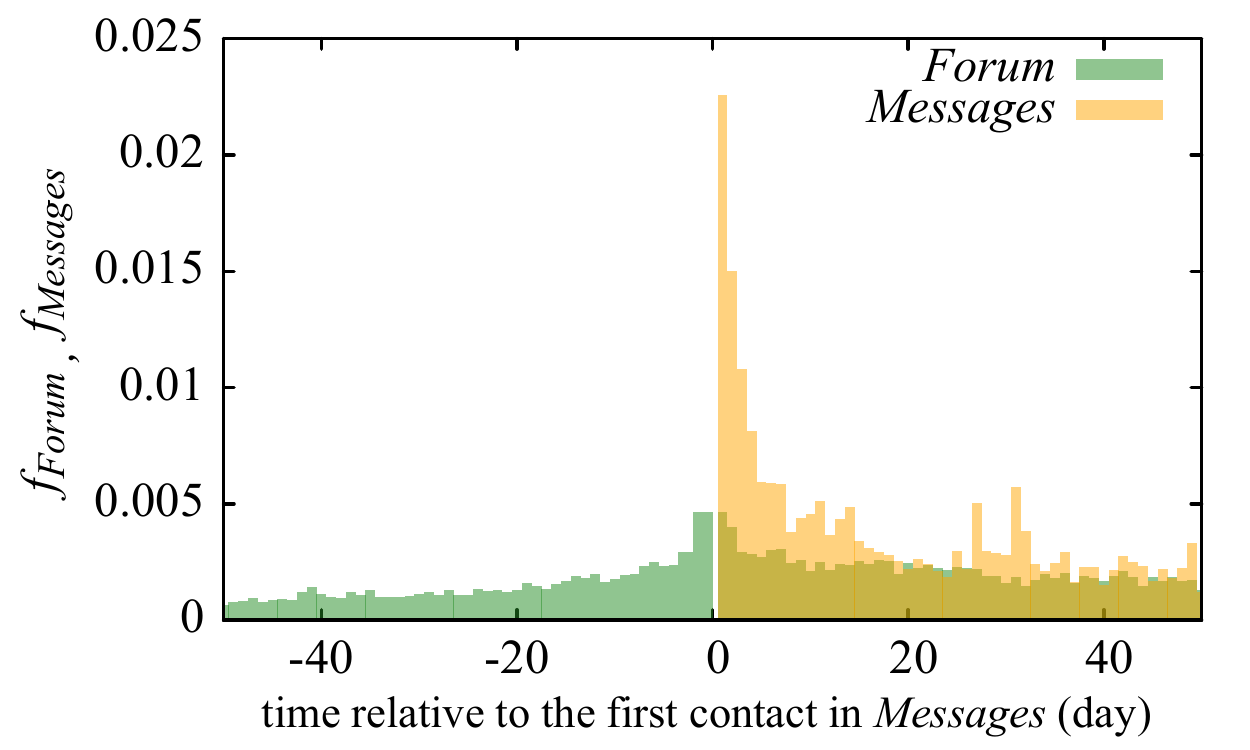}
\caption{\label{fig:pair_transfer}\textbf{ Frequency of contacts relative to the first contact in \emph{Messages} for dual-channel pairs.} The diagram shows that pairs who contact in \emph{Messages} at $t=0$, later, shift the majority of their communications to \emph{Messages}. We do not observe any dramatic changes in the \emph{Forum} contact frequency. The frequency at $t=0$ is not shown. At $t=0$ $f_\mathrm{Forum} = 0.02$ and $f_\mathit{Messages} = 0.17$. The relatively high value of \emph{Forum} frequency at $t=0$ suggests that initially the \emph{Forum} communication triggers shift of the communication to the \emph{Messages}. The bin is measured by day. }
\end{figure}

As we mentioned earlier, there are 7,183 dual-channel pairs who use both \emph{Forum} and \emph{Messages} for their communication.  Here we focus on the 4,812 of these pairs that start their communication initially in \emph{Forum}.  In Fig.~\ref{fig:pair_transfer}, we show the frequency of contacts between those pairs before and after their first contact in \emph{Messages}. At $t=0$ the pair sent their first contact in \emph{Messages} and other contact times are calculated relative to the first contact in \emph{Messages}. The figure shows that prior to the first contact in \emph{Messages}, the frequency of contacts increases in \emph{Forum}. However, after the first contact in \emph{Messages}, the majority of the contacts are shifted to the \emph{Messages} channel. The relatively high value of \emph{Forum} frequency at $t=0$ ($f_\mathit{Forum} = 0.02$) suggests that initially the \emph{Forum} communication triggers shift of the communication to the \emph{Messages}. For the remaining dual-channel pairs who start their first contact in \emph{Messages}, we do not observe any significant differences in their contacts rate before and after the first contact in \emph{Forum}.

 In the following section we investigate whether the differences between networks we observed in individuals' behavior are also expressed structural differences at the community level.

\subsection{Structural comparison of the two communication channels}
\paragraph{Reciprocity}
Reciprocity measures the ratio of bidirectional edges in networks ~\cite{davis_reciprocity}. Reciprocity captures the tendency for communication to go back and forth as in a dialogue. To be able to identify the significance of the reciprocity compare to a random network, we use the definition based on Ref.~{\cite{garlaschelli2004patterns}. Assume the edge density of the network is $ \langle L \rangle=L/N(N-1)$, where $L$ is total number of edges and $N$ is total number of vertices. The reciprocity is   
\begin{equation}
\rho = \frac{r-\langle L \rangle}{1-\langle L \rangle}
\end{equation}
where $r$ is the fraction of the reciprocated edges, $\frac{L_{\leftrightarrow}}{L}$.} The values are reported in Table~\ref{tab:network_properties}. \emph{Messages} showed a higher degree of reciprocity than \emph{Forum}, which indicates that reciprocal interactions play a more important role in structuring the personal communication in online space. This confirms the insight~\cite{harasim_global_networks} that without reciprocity, online communities fail and going online becomes an unpleasant routine task.

\paragraph{Assortativity}
Assortativity measures how individuals with a certain degree are connected to other individuals~\cite{newman_assortativity}. We compared assortativity values against those obtained by a null model  generated by reshuffling the edges between vertices while maintaining their degree distribution. This reshuffling method generates networks with the same size and degree sequence but destroys all other structure. By comparing to the null model, we can understand how much of the structure of the network comes from other factors than the degree. For the assortativity measure we used undirected networks.
Results from Table~\ref{tab:network_properties} show that there is a weak tendency in the \emph{Forum} and \emph{Messages} networks to be more assortative compared to the null model. Hence, the degree distribution can generate similar results with respect to the empirical data. The disassortativity has also been observed in other online social networks \cite{pok, Ahn_online_ntw,Mislove:2007,disassortative}. 

\paragraph{Transitivity}
Transitivity measures the fraction of existing triangles divided by all possible triangles in the network~\cite{holland_transitivity}. For the transitivity measure, we used undirected networks. Transitivity captures the number of triads that are connected and form triadic social interactions. Our goal is to investigate whether triadic social interactions are significantly more common than what would be expected just by the distribution of the degree. Furthermore, we compared the transitivity of the \emph{Messages} and \emph{Forum} networks against that yielded by randomized null models of the networks. The randomized null model is generated as mentioned in the previous section. Results from Table~\ref{tab:network_properties} show that for both networks, transitivity was slightly higher than that of the corresponding randomized model. Interestingly, we observe higher transitivity in \emph{Forum} compare to the \emph{Messages}. The characteristic of the online forum to enhance open communications and discussions, enforces conversational cliques that do not exist in \emph{Messages} \cite{dunbar_conversational_size}. 

The reason is that offline social networks have many triangles \cite{watts1998collective}. There are two mechanisms for that. First, people belong to tightly knit communities (of family, colleagues, etc.). A tightly connected subnetwork would (per definition) have a large fraction of edges and thus have many triangles.  The second mechanism is that people introduce their friends to each other, which results in the formation of a triangle.

\begin{table}
\caption{\textbf{Network properties of
the \emph{Forum} and \emph{Messages} data.} We compare the results for the empirical networks  with a randomized model (shown in brackets). For all cases $P\leq 0.01$}
\label{tab:network_properties}
\begin{tabular}{llll}
\hline \hline
Network &  Assortativity &  Transitivity   &  Reciprocity\\
\hline 
\emph{Forum} & $-0.286$ {[}$-0.315${]}  &  $0.26$ {[}$0.21${]} & 0.59\\
\emph{Messages} & $-0.037$ {[}$-0.044${]}  &  $0.05$ {[}$0.00${]} & 0.65\\
\hline\hline
\end{tabular}\end{table}

\subsection{Social structure of the community interaction}
\paragraph{Communication motifs}
Triadic social interactions can be classified according to their stability~\cite{antal_balance,wasserman_triangle,szell_online_game}. Structural balance theory provides a way of describing triangle stability by categorizing the edges between vertices as positive (friend) or negative (enemy) social ties. If the ties that make up a triangle are multiplied and result in a non-negative value, the triadic interaction is classified as balanced. If there is an odd number of negative ties, then the interaction is considered unbalanced. The two types of interaction we study do not have any direct relation to friendship or animosity, thus our analysis will not be a test of structural balance theory, or stability of triangles.  Instead, the multiplex nature of this dataset did allow us to investigate the frequency of various communicational motifs (i.e., various configurations of \emph{Forum} or \emph{Messages} ties). We assigned labels for ties that can be either \emph{Messages} (M) or \emph{Forum} (F) depending on  which channel that was used more by the pair. Based on these labels in total there are 90,148 M-edges and  119,694 F-edges. Then we measured different number of social motifs between the two networks. 

\begin{figure}
\includegraphics[width=.9\linewidth]{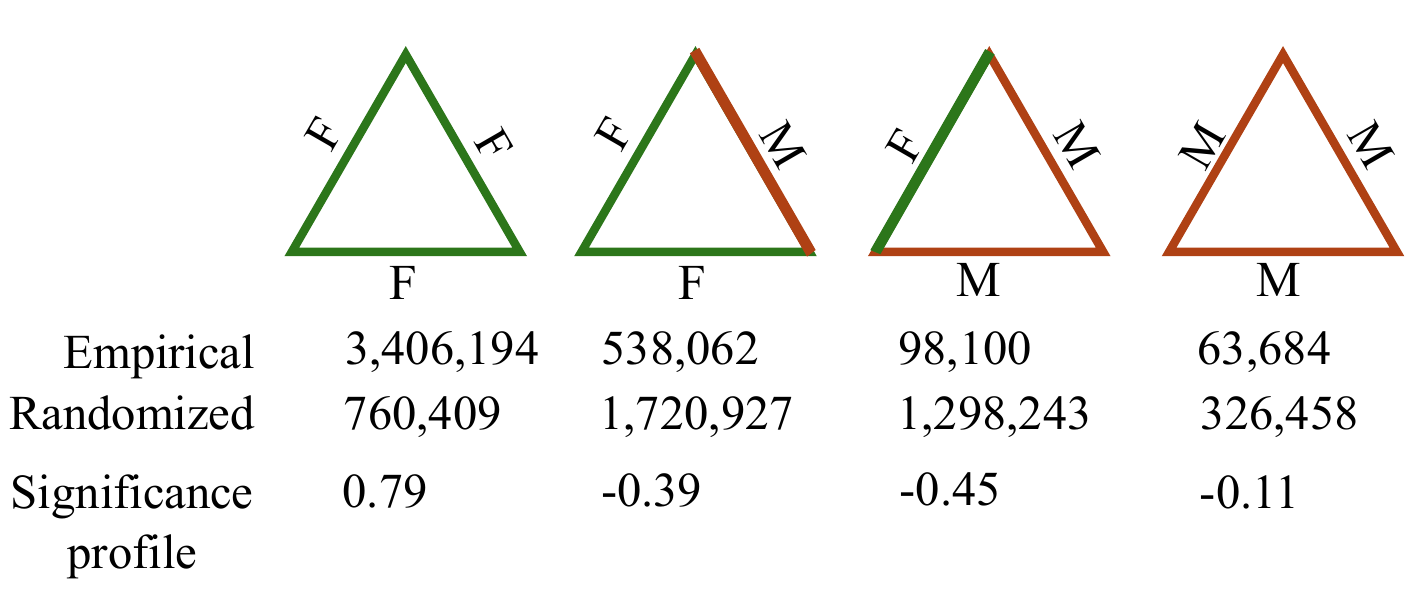}
\caption{\label{fig:triangles}\textbf{Communication motifs in \emph{Forum}
and \emph{Messages} networks.} Edges between pairs are labeled as \emph{Forum} or \emph{Messages}. The label indicates that which communication channels the pairs has used more frequently. Results are compared to randomized model by keeping the edges and reshuffling the labels. }
\end{figure}

In principle, a network can be composed of a combination of four kinds of triangular motifs: 1) M-M-M, 2) F-F-F, 3) M-F-F, and 4) M-M-F. 
We measured the frequency of these four kinds of triangles in the total dataset and compared it against the frequencies obtained by a null model where the edge labels, M and F, are distributed merely based on the expected ratio of the label occurring. The comparison of results with the null model shows the significant motifs that can not be expected only by random. Based on the empirical observation, we define the frequency of occurring F edges, $f$ that is $0.57$ and consequently the frequency of M edges, $m$, is $0.43 $. Therefore the expected fraction of F-F-F triangles is $f^3$ (equalling 760,409 in our case), the expected fraction of F-F-M triangles is $3f^2(1-f)$ (1,720,927 in our case), etc. Furthermore, We calculate the Z-score and significant profile (SP) for each motif \cite{Milo}. The significant profile is a useful tool because the Z-score is sensitive to the network size. If the Z-score for each motif $i$ is denoted by $Z_{i}$, the significant profile of the motif, $SP_{i}$ is defined:

\begin{equation}
SP_{i} = \frac{Z_{i}}{\sqrt{\sum_{i}Z_{i}^2} }
\end{equation}

The Results are shown in  Fig.~\ref{fig:triangles}. The only over-represented motif in the community was the F-F-F triadic relationship.  This suggests that the building blocks of the online community is originated by contacts in \emph{Forum}. Moreover, it can be interpreted as homophily---the tendency to get acquainted to similar others~\cite{mcpherson_homophily}. The most under-represented motif was the M-M-F combination. Assuming that interactions in \emph{Messages} are related to stronger social ties than \emph{Forum} contacts, the prevalence of message-message-forum triangle can tell us something about the type of relationships that emerge in the community. For instance, if A-B and A-C are both strong ties, it is more likely that B-C will also be a strong tie.

\paragraph{Network similarity}

\begin{figure}
\includegraphics[width=1\linewidth]{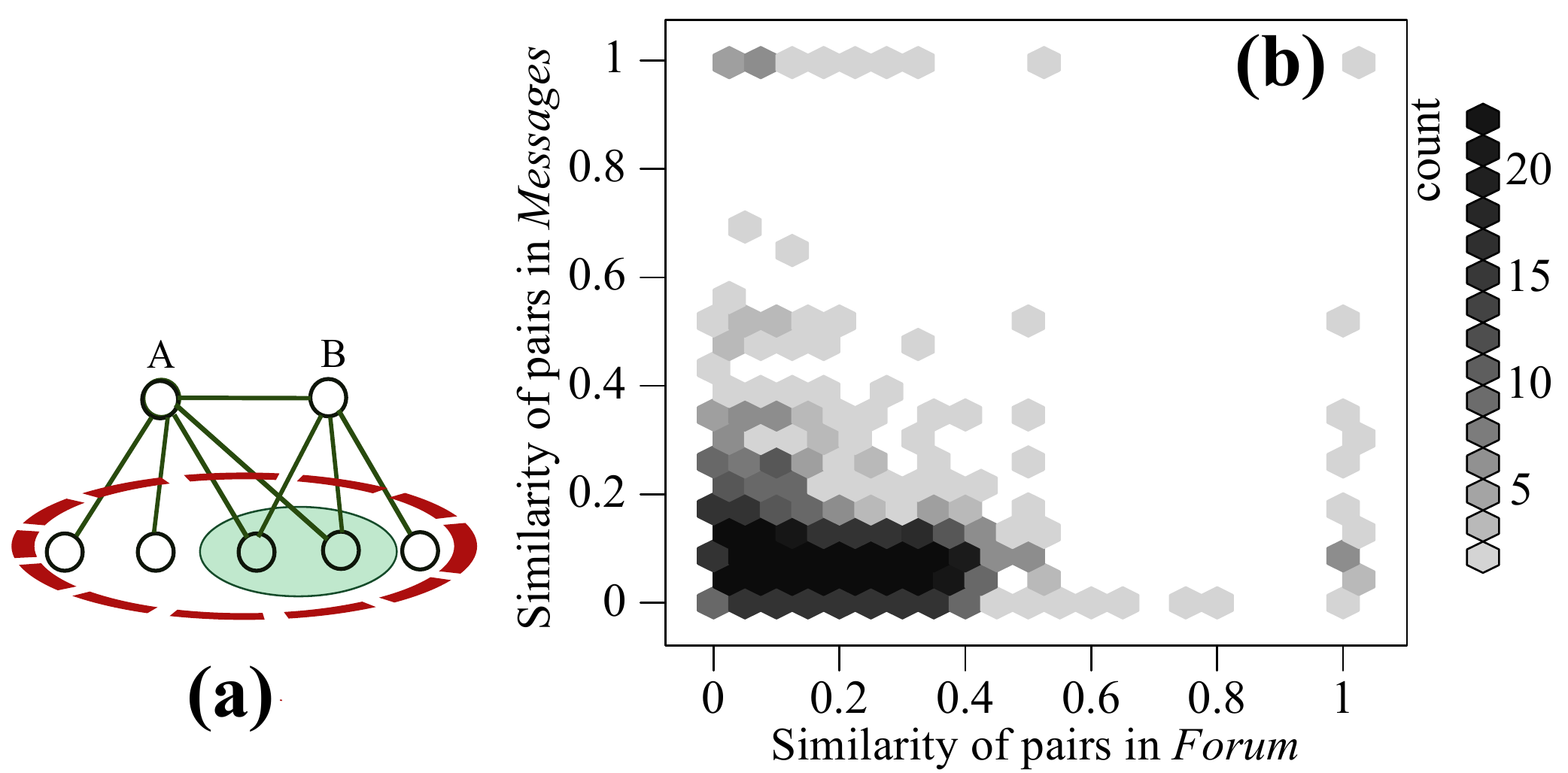}
\caption{ \textbf{Jaccard similarity of pairs comparing the similarity in
\emph{Forum} and \emph{Messages} network.} Figure \emph{a} illustrates the Jaccard similarity index measure. In this example, pair AB has two common neighbours, and therefore their Jaccard similarity is $2/5$. Figure \emph{b} displays Jaccard similarity of pairs in \emph{Forum} and \emph{Messages}.  White spaces indicate absence of a pair with this combination of similarity values. The gray-scale bar shows the number of pairs with a certain similarity. }
\label{fig:similarity_fm}
\end{figure}

We used the Jaccard similarity index to evaluate the similarity between pairs in terms of their neighbors. According to the similarity measure, user A and user B in the community are similar if they share a larger number of common neighbors, see Fig.~\ref{fig:similarity_fm}. If user A has the neighborhood $\Gamma(A)$ and user B has the neighborhood $\Gamma(B)$, the Jaccard similarity of A and B, $J(A,B)$, is:
\begin{equation}
J (A,B) = \frac{|\Gamma(A)\cap \Gamma(B)|}{|\Gamma(A) \cup \Gamma(B)|}
\end{equation}

Here we report the similarity of pair between \emph{Forum} and \emph{Messages} respectively. For other choices of similarity measures please see Ref.~\cite{zhou_similarity}. The majority of the pairs had a Jaccard index of zero in the networks, meaning that no common neighbours have been found between the pair. For those pairs with non-zero similarity, Fig.~\ref{fig:similarity_fm} shows the scatter plot of the Jaccard indices of similarity in the two modes of communication respectively. We see that there are many pairs with low similarity in both \emph{Forum} and \emph{Messages}, but if they have a high similarity in one of the networks---horizontal line in the top left in \emph{Messages} and vertical line in the bottom right in \emph{Forum}---they typically do not have high similarity in the other network. In the Fig.~\ref{fig:similarity_fm} we plot a scatter plot of the similarities between vertex pairs. What we find that there is no distinct correlation of similarities of pairs between the two communication channels.

\paragraph{Appropriation of technology}

\begin{figure}
\includegraphics[width=.8\linewidth]{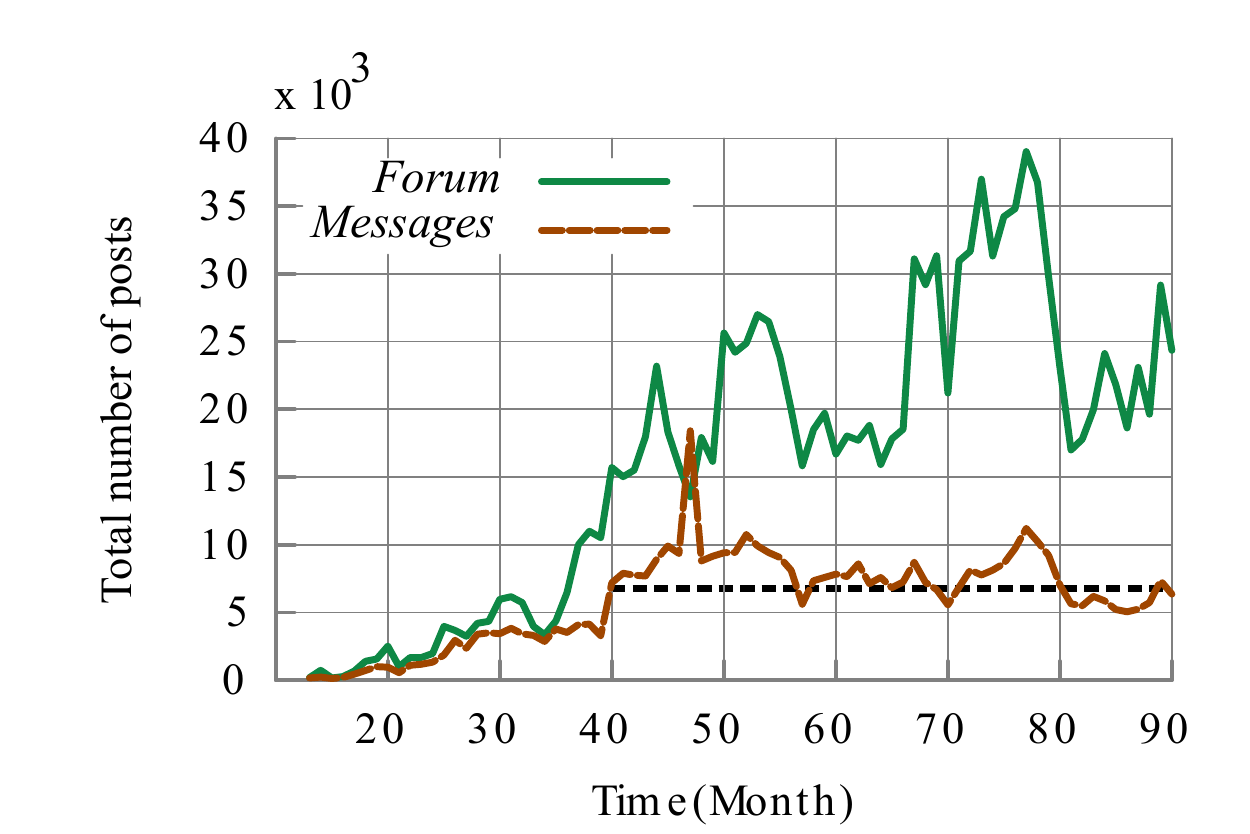}
\caption{ \textbf{Total number of forum posts and private messages that were exchanged in the platform as a function of time.}  }
\label{fig:FT_popularity}
\end{figure}

Users assign subjective meaning to the technology they consume, and their usage of technology depends on the meaning \cite{mackay1992extending}. For example, we see that not so many users use both channels, while many users clearly prefer one communication channel over the other (\emph{Forum} or \emph{Messages}). As Douglas and Baron put it ``Goods are neutral, their uses are social, they can be used as fences or bridges''~\cite{douglas1996world}. The functionality of the technology is two-sided: one that has been developed by designers and their intentions and one that has been shaped by users. The current dataset allows us to observe the activity level of users from the very beginning of the creation of the website. The website is initially designed to provide a space for users to discuss and share their interests in movie-related topics. Therefore, we can assume that activity in \emph{Messages} is mainly driven by users. 
Here, we simply measure the total forum posts and messages that were exchanged over time in the whole community (see Fig.~\ref{fig:FT_popularity}). Interestingly, from the begining of creation of the website, the level of users engagements are similar in both platforms. However, once the community is shaped, the level of activity in \emph{Messages} is stable whereas activity level in \emph{Forum} is influenced by externalities such as trendy movie related topics.

\section{Discussion}
We studied communicational aspects of an online community in terms of open (\emph{Forum}) versus direct (\emph{Messages}) communication within the same population. 
 At the individual level, we find that people are using the two networks in different ways which can be understood in the light of the means of communication.  For example, the level of social involvement is higher in \emph{Forum}, whereas \emph{Messages} platform is designed for reciprocity and deeper social investment. The distribution of in-degree and out-degree is broader in \emph{Forum}. The edge weight distribution displays broader tail in \emph{Messages} which indicates broader range of pair communicational exchange. 
 
Furthermore, we compare the interevent of activity for individuals in the two channels of communication. That allow us to investigate how the means of communication drives time allocations among individuals. The interevent times show that while a user logged in (i.e., for small time windows), she is slightly faster in replying to personal emails than \emph{Forum} posts. Within 24 hours, the interevent times become broader in \emph{Forum}, indicating the engagement in discussions needs to be quick. After 24 hours from the last login, the user is more likely to respond \emph{Messages} than \emph{Forum} communication. These observations can also be explained by \emph{Messages} supporting stronger social ties and \emph{Forum} facilitating creation of new ties.

We introduced a measure called sociality that captures the individuals capability of creating ties. It turns out that, not surprisingly, sociality is more broadly distributed in \emph{Forum}. We also defined a measure called interactivity that captures the overall activity of a user. The interactivity has broader range in \emph{Forum}, yet there is an upper limit of interactivity in both platforms. Furthermore, we measured the time interval that individuals take to create a new edge. Interestingly, we observe that, as opposed to \emph{Forum}, waiting time to create a new edge in \emph{Messages} increases with the number of edges. In other words, creating edges in \emph{Messages} involves a greater social effort. For those pairs who use both means of communication, once a \emph{Messages} edge has been created, the majority of their communication transferred to direct \emph{Messages}.

The detailed analysis of individual behavior leads to understanding the meso scale structure of the system that were described above. Despite the differences in the structure of the two networks in the meso scale, they exhibit similar structure in macro scale in terms of assortativity and transitivity. Furthermore, we looked at triangle motifs as a larger unit of network organization. It turns out that the forum-forum-forum triangle is the only over-represented motif. This reflects the fact that the building blocks of the community are created in \emph{Forum}.  In terms of the pair similarity, we did not find a significant correlation between similarity of a pair in \emph{Forum} and \emph{Messges}.

In sum, our results show that despite similarities in the macro scale structure of the two networks,  in the meso scale, the structure differs depending on the context in an online community. This suggests that electronic communication has some universally true aspects (broad degree and interevent time distributions, high transitivity and low assortativity), but also some context dependent properties that are related to a particular mode of communication. One future research direction is to model realistic social interactions based on the means of communications. Furthermore, the findings from this analysis can be used to link prediction on the multiplex networks \cite{Lu_link_prediction} and to study how various channels of communication facilitate information spreading in online space.

\section*{Acknowledgement}
The authors thank Magnus Hoem and Jonny Lundell for help with the data. We also thank Alcides Viamontes Esquivel, Barbara Pabjan, Andrea Lancichinetti, Daniel Stouffer, Matthias Raddant, Ann Samoilenko and Renaud Lambiotte for helpful discussions. We are grateful for anonymous reviewers for insightful comments and suggestions. This research was supported by Basic Science Research Program through the National Research Foundation of Korea (NRF) funded by the Ministry of Education (2013R1A1A2011947) and the Swedish Research Council.

\end{document}